\documentclass{optica-article}

\journal{opticajournal} 

\articletype{Research Article}

\usepackage{lineno}
\usepackage{subfigure}

\begin{document}

\title{Amplification and Excitation of Surface Plasmon Polaritons via Four-Wave Mixing Process}

\author{Andleeb Zahra,\authormark{1}, Muqaddar Abbas, \authormark{2} 
Rahmatullah,\authormark{1},\authormark{*}}

\address{\authormark{1} Quantum Optics Lab, Department of Physics, COMSATS University Islamabad, Pakistan.}
\address{\authormark{2}Ministry of Education Key Laboratory for Nonequilibrium Synthesis and Modulation of Condensed Matter, Shaanxi Province Key Laboratory of Quantum
Information and Quantum Optoelectronic Devices, School of Physics, Xi’an Jiaotong University, Xi’an 710049, China.}

\email{\authormark{*}rahmatktk@comsats.edu.pk} 


\begin{abstract*} 
We suggest a scheme for the excitation and amplification of surface plasmon polaritons
(SPPs) along the interface between metal and semiconductor quantum well (SQW), employing a four-wave mixing (FWM) process. The SQW consists of four-level asymmetric double quantum wells that exhibit quantum interference effects, which leads to the coupler-free excitation of SPPs. In our proposed system, the inherent losses of SPPs are compensated by introducing gain through the FWM process. This results in a significant enhancement in the propagation length and large penetration depth of SPPs. We further analyze the effect of gain on the long-range and short-range SPPs and observe that the propagation distance and lifetime of both types of SPPs are enhanced.

\end{abstract*}

\section{Introduction}
Surface Plasmon Polaritons (SPPs) are surface electromagnetic waves that arise from the interaction of the incident light with the free electrons oscillations in metal, and propagate along the metal-dielectric interface within the frequency ranges from visible to infrared (IR). These surface waves have evanescent-like behavior in the plane normal to the interface. The history of SPPs goes back hundreds of years \cite{stern, PhysRevB.10.3342}, with the initial observation credited to Robert Wood in 1902 \cite{wood1902xlii}. However, intensive research has been started after the investigation of the plasmonic properties of the silver and gold nanoparticles \cite{kreibig1970surface}. SPPs have the peculiar property to confine the light to the nanoscale by providing relaxation from the classical diffraction limit. It marks them special for the fabrication of on-chip integrated plasmonic devices \cite{dong2015recent}. So far, SPPs are widely used in biomedical and chemical sensors, photo-lithography \cite{Phys.Rev.A95.053850}, nonlinear nano-scale photonics \cite{shi2021nonlinear}. The propagating nature of SPPs is also exploited for information transfer including the plasmon-assisted transmission of entangled photons \cite{altewischer2002plasmon,fasel2005energy}, and the SPP-mediated quantum teleportation \cite{jiang2020quantum}. However, there is a characteristic trade-off between the increased confinement of SPPs at the interface and their large propagation distance. Typically, a single interface structure with two semi-infinite media sustains only a single mode of SPPs. On the contrary, the dispersion relation for propagating SPPs in thin films surrounded by two media with similar or different refractive indices splits into two branches \cite{agranovich2012surface,burke1986surface,economou1969surface,zayats2005nano}, indicating the two distinct (symmetric and anti-symmetric) modes of SPPs. The existence of these modes on unsupported thin metal films is observed theoretically and also experimentally verified \cite{kuwamura1983experimental}. The former mode has a comparatively large propagation length than the single interface SPPs, and these are referred to as long-range SPPs (LR-SPP) \cite{sarid1981long,berini2009long}. Whereas, the short-range SPPs (SR-SPPs) show increased confinement to the metal film. Since SPPs are sensitive to surface conditions, as a result, they decay rapidly and have small propagation distances, which limit their practical applications. Earlier on, structured metallic films \cite{kocabas2009slowing} or grating \cite{sondergaard2006theoretical} are used in slowing down the SPPs and enhancing their propagation length. A few years later, electromagnetically-induced transparency (EIT) is harnessed to slow down the SPPs at the interface of dielectric and active negative index meta-material \cite{kamli2008coherent}.

The excitation of SPPs requires energy and momentum conservation. The dispersion relation of SPPs is given by; $k_{SPP}=\frac{\omega}{c}\sqrt{\frac{\epsilon_m\epsilon_d}{\epsilon_m+\epsilon_d}}$ \cite{maier2007plasmonics}(where $\epsilon_m$, $\epsilon_d$ are the relative permittivities of the metal and dielectric medium, respectively). This dispersion relation shows that the momentum of an SPP is typically much smaller than the momentum of the incident light. This momentum mismatch prevents the direct excitation of SPPs from light, which is a challenge for the practical implementation of SPPs. Several schemes are thus proposed to circumvent all the challenges associated with the SPPs. Initially, Otto and Kretschmann proposed a prism (a tool to increase the momentum of light) that exhibits frustrated total internal reflection (FTIR). However, geometrical challenges are inherent to these methods. In 2015, EIT is characterized for the very first time to observe the coupler-free transition of SPPs from light \cite{PhysRevA.91.013817}. It is investigated that the direct excitation of SPPs is possible only if the dielectric medium is EIT-based and has the real part of its permittivity less than one. Moreover, nonlinear \cite{PhysRevLett.101.056802}, and free space excitation of SPPs \cite{PhysRevLett.103.266802} is demonstrated by the means of the four-wave mixing (FWM) process. However, losses present in the systems are not negligible, which leads to weak SPR resonance. Using a gain medium is one of the possible solutions for the compensation of losses. Many coupler-based schemes are already proposed in this regard
\cite{avrutsky2004surface,nezhad2004gain,seidel2005stimulated}, and quantum wells are found to be good candidates as the gain medium.

In general, SQWs have discrete energy levels and atomic-vapors-like optical properties. During the past few decades, quantum coherence phenomena including lasing without inversion \cite{Imamoglu:94,PhysRevB.59.12212,frogley2006gain}, coherent population trapping \cite{PhysRevB.72.085323}, EIT \cite{PhysRevLett.89.186401,silvestri2002electromagnetically,wang2003electromagnetically}, enhancement of refractive index \cite{PhysRevB.62.15386}, and slow light \cite{ku2004slow} are intensively studied both theoretically and experimentally in SQWs. In addition, the FWM process is explored in several schemes of SQWs \cite{hao2008efficient,evangelou2011pulsed,kosionis2013transient,liu2014enhanced}. The FWM process, a flourishing nonlinear process, is the subject of growing research for the past few years, due to its potential applications in quantum information \cite{PhysRevLett.91.123602}, nonlinear optics \cite{PhysRevLett.91.093601} and spectroscopy \cite{PhysRevLett.82.4611}. 

In this article, we explore the amplification and quantum-coherence-driven excitation of SPPs along the metal-SQW interface via FWM process. Furthermore, we introduce a thin metal film to exploit the propagation of long-range (LR) and short-range (SR) SPPs. Our proposed scheme is based on a three-layer structure, where the bottom layer is the asymmetric SQW that exhibits an EIT-based FWM process. In our system, not only does the SQW as a gain medium portrays gain, but also the generated fourth wave through the FWM process provides sufficient compensation for the losses, leading to the resonant excitation and amplification of SPPs with their increased propagation distance along the metal-SQW interface. In addition, by means of the FWM process, LR-SPPs propagate for a longer period to a larger distance. Also, our system is more feasible than the atomic counterpart as the quantum interference effects and other parameters are easily tunable.

\section{Model and Equations}
\begin{figure}[t]
    \centering
    \includegraphics[width=12cm]{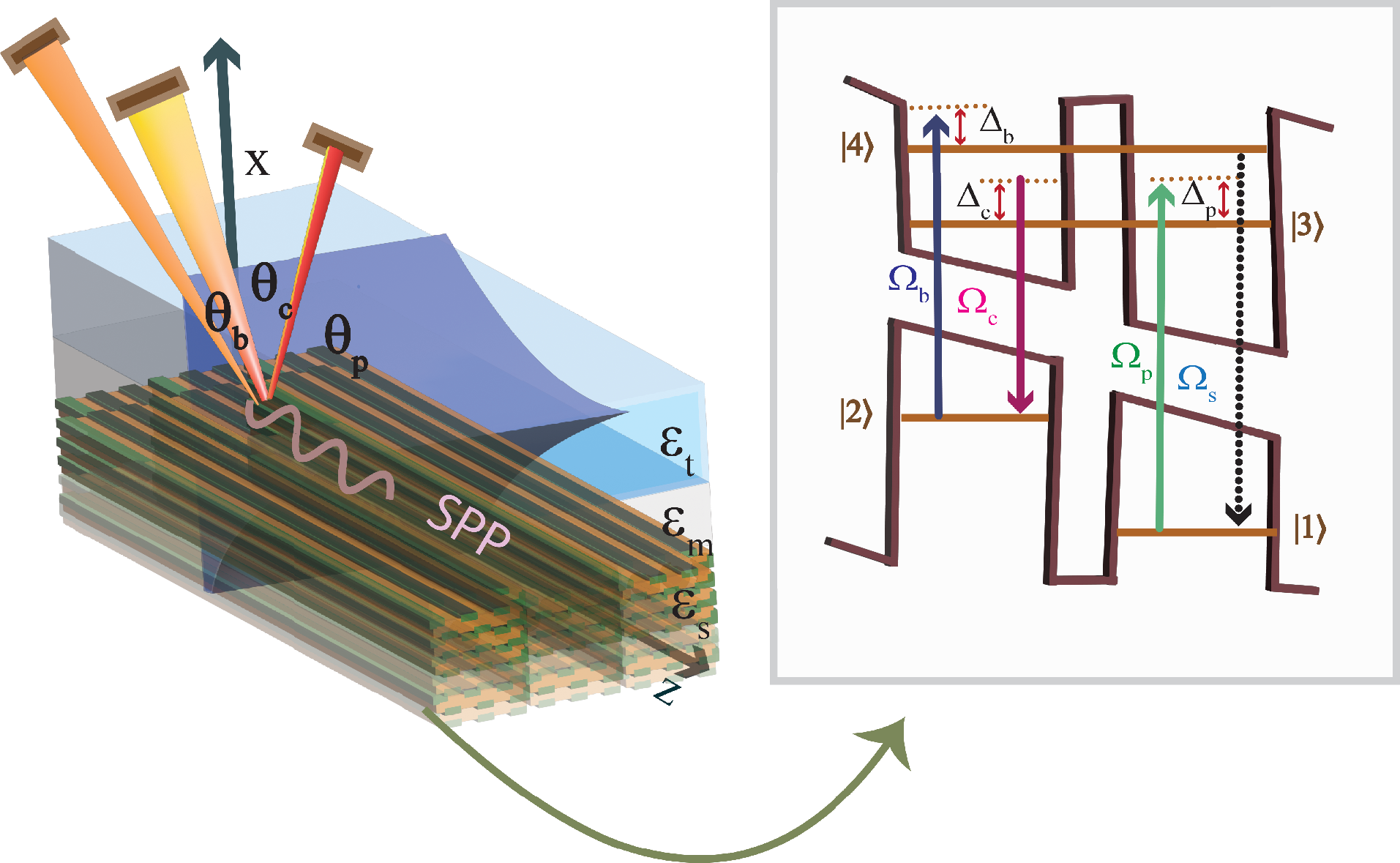}
    \caption{(Color online) Schematic of SPR system with SQW as a quantum medium. The system consists of a three-layer structure with a transparent top layer (blue), a metal film (gray), and a bottom layer of SQW (green), comprised of four-level asymmetric double quantum wells. The inset shows the energy level configuration of the SQWs, enabling the EIT-based FWM process.}
    \label{model}
\end{figure}

The schematic to study the SPPs for coupler-free SPR with relatively enhanced propagation length due to the FWM process is shown in Fig. \ref{model}. The system consists of a three-layer structure, where a metal film separates the transparent top layer (vacuum/air) from an EIT-based bottom layer that is composed of four-level asymmetric double SQWs. $\epsilon_t, \epsilon_m$ and $\epsilon_s$ are the relative permittivities of the top, middle and bottom layers, respectively. To initiate the excitation of the SPPs at the metal-SQWs interface, three electromagnetic fields including a weak probe field, a strong control field, and a pump field are incident with angles $\theta_p$, $\theta_c$, and $\theta_b$, respectively, at the top-layer-metal interface. Owing to EIT conditions, the permittivity of SQWs is coherently controllable and can lead to the resonant excitation of SPPs with reduced propagation losses. Also, our structure supports two SPPs modes (symmetric (Long-range) and anti-symmetric (Short-range)) with different propagation lengths in the limit of very thin metallic films (where $q<<\lambda_0$).  
\par Based on the recent experimental condition \cite{roskos1992coherent}, we consider asymmetric double SQW  following a four-level configuration \cite{kang2016enhanced, luo2011nonlinear}. Fig. \ref{model} illustrates the structure of an asymmetric double SQW that can be fabricated by using 10 pairs of wide wells (WW) each with a thickness of 51-monolayer ($145$ Angstrom) and $35$-monolayer ($100$ Angstrom) thick narrow well. Which are then separated by a thin AlGaAs barrier having a thickness of $9$-monolayer ($25$ Angstrom). As shown in Fig. \ref{model}, the valence band consists of levels $|1\rangle$ and $|2\rangle$ which are localized hole states. The conduction band consists of levels $|3\rangle$ and $|4\rangle$, and these are delocalized bonding and anti-bonding electronic states, respectively having energy difference $\omega_s$. These electronic states are generated due to tunneling (through a thin barrier) in two quantum wells. Two coherent fields i.e., probe and control fields, govern the transitions among the electronic and hole states and induce EIT. Probe field with amplitude $\mathcal{E}_p$ and Rabi frequency $\Omega_p$ initiates the transition $|3\rangle \leftrightarrow |1\rangle$ with transition frequency $\omega_{31}$. Whereas, a control field of amplitude $\mathcal{E}_c$ and frequency $\Omega_c$ is coupled to the transition from electronic state $|3\rangle$ to hole state $|2\rangle$ having transition frequency $\omega_{32}$. To drive intersubband transition $|4\rangle \leftrightarrow |2\rangle$, with natural frequency $\omega_{42}$, another field named pump field having amplitude $\mathcal{E}_b$ and Rabi frequency $\Omega_b$, is applied. These three fields then induce a FWM process $|1\rangle \rightarrow |3\rangle \rightarrow |2\rangle \rightarrow |4\rangle \rightarrow |1\rangle$, eventually, a coherent radiation field with Rabi frequency $\Omega_s$ is generated.

Under the rotating-wave approximation, the interaction picture Hamiltonian is given by ( $\hbar=1$)
\begin{eqnarray}
 H_I&=&-\Delta_p|3\rangle \langle 3| -(\Delta_p-\Delta_c)|2\rangle \langle 2| -(\Delta_p-\Delta_c  +\Delta_b)|4\rangle \langle 4|  -(\Omega_p e^{i\mathbf{k_p.r}}|3\rangle \langle 1|+ \nonumber \\&& \Omega_c e^{i\mathbf{k_c.r}}|3\rangle \langle 2| +\Omega_b e^{i\mathbf{k_b.r}}|4\rangle \langle 2| 
 +\Omega_s e^{i\mathbf{k_s.r}}|4\rangle \langle 1|+H.c),
\end{eqnarray}
where $\Delta_p=\omega_p-\omega_{31}$, $\Delta_c=\omega_c-\omega_{32}$ and $\Delta_b=\omega_b-\omega_{42}$ are the detunings of probe, control, and pump fields, respectively and $\omega_p$, $\omega_c$, and $\omega_b$ are their corresponding angular frequencies.

Then, by using the linear Schrodinger wave equation, $i\frac{\partial |\psi\rangle}{\partial t}=H_I|\psi\rangle$, with $|\psi\rangle =A_1+A_2 e^{i(\mathbf{k_p}-\mathbf{k_c}).\mathbf{r}}+A_3 e^{i\mathbf{k_p.r}}+A_4 e^{i(\mathbf{k_p}-\mathbf{k_c}+\mathbf{k_b}).\mathbf{r}}$, we get the following  equations of motion for probability amplitudes
\begin{equation}
\frac{\partial A_1}{\partial t}=i \Omega_p^* A_3 + i \Omega_s^* e^{i\delta\mathbf{k.r}} A_4,
\end{equation}
\begin{equation}
\frac{\partial A_2}{\partial t}=i[(\Delta_p - \Delta_c)+i \gamma_2]A_2 +i \Omega_c^* A_3 + i \Omega_b^* A_4,
\end{equation}
\begin{equation}
\frac{\partial A_3}{\partial t}=i (\Delta_p+i \gamma_3)A_3 + i \Omega_p A_1 +i \Omega_c A_2 + \kappa A_4,
\end{equation}
\begin{equation}
\frac{\partial A_4}{\partial t}= i [(\Delta_p -\Delta_c + \Delta_b)+i \gamma_4]A_4 + i \Omega_s  e^{-i\delta\mathbf{k.r}} A_1 + i \Omega_b A_2 + \kappa A_3,
\end{equation}
where, $\delta \mathbf{k}=\mathbf{k}_p+\mathbf{k}_b-\mathbf{k}_c-\mathbf{k}_s$, is the phase mismatching factor. Under the steady-state and the phase matching condition $\delta \mathbf{k}=0$, the optical susceptibility of the SQW is given by 
\begin{equation}
\chi=\frac{N|\mu_{13}|^2}{\epsilon_0 \hbar \Omega_p}\times (A_3 A_1^\ast),
\end{equation}
after substitution, we get
\begin{equation}
\chi=\frac{N|\mu_{13}|^2}{\epsilon_0 \hbar \Omega_p}\times \frac{(\Omega_b^2-d_2 d_4)\Omega_p-(\Omega_c^\ast \Omega_b + d_2i\kappa)\Omega_s}{(-i\kappa(\Omega_c^* \Omega_b  +\Omega_c \Omega_b^*)-(d_4 \Omega_c^2 + d_3\Omega_b^2) + d_2(d_3 d_4 +\kappa^2))},
\end{equation}
where $N$ is the number density of electrons in the conduction band of SQW. And,  $d_2=(\Delta_p-\Delta_c)+i\gamma_2$, $d_3=\Delta_p+i\gamma_3$, and $d_4=(\Delta_p -\Delta_c+\Delta_b) +i\gamma_4$ with $\gamma_j=\gamma_{jl}+\gamma_{jd}$ (j=2, 3, 4) that describes the total decay rate of level $|j\rangle$, where $\gamma_{jl}$ is the decay rate due to longitudinal-optical (LO) phonon emission, and $\gamma_{jd}$ is dephasing decay rate between levels $|i\rangle \leftrightarrow |j\rangle$. The parameter $\kappa=\sqrt{\gamma_{3l}\gamma_{4l}}$ defines the cross-coupling of the state $|3\rangle$ and $|4\rangle$, and it gives rise to interference between bonding and anti-bonding states\cite{liu2014enhanced}.

The permittivity of the SQW related to the susceptibility is written as \cite{aspnes1982local,crenshaw1996local}
\begin{equation}\label{permittivity of semiconductor}
    \epsilon_s = 1+\frac{\chi}{1-\frac{1}{3}\chi}.
\end{equation}
Typically, the excitation of SPPs is explored through the reflection of the probe beam, and the transmission gives the electromagnetic field enhancement within the small volume. So, by Fresnel equations, the reflection coefficient for our proposed scheme is given by \cite{reather1988surface} 
\begin{equation}\label{ reflection coefficient}
    r_{tms} = \frac{r_{tm} + r_{ms} exp(2i k_{mx} q) }{1+r_{tm} r_{ms} exp(2i k_{mx} q) },
\end{equation}
where q is the thickness of the metallic film and $r_{ij}$ is the reflection coefficients for a single interface. In case of TM-polarized light, it is expressed as
\begin{equation}\label{rtm}
    r_{ij} = \frac{\epsilon_j k_{ix} - \epsilon_i k_{jx}}{\epsilon_j k_{ix} + \epsilon_i k_{jx}},
\end{equation}
where $(i)j = t,m,s$ represent top-layer, metal film, and semiconductor medium, respectively, and $k_{jx}$ is the normal wave vector which is defined as  
\begin{equation}\label{kjx}
    k_{(i)jx} = \sqrt{k_0^2 \epsilon_{(i)j} - k_z^2},
\end{equation}
where $k_z$ is the in-plane component of the wave vector that depends on the free space wave vector of incident probe beam $k_0$, top-medium refractive index $n_t$ and probe field angle of incidence $\theta_p$ that can be defined as
\begin{equation}\label{kz}
   k_z = k_0 n_t \text{sin}\theta_p.
\end{equation}
Likewise, the transmission coefficient is given by
\begin{equation}\label{transmission coefficient}
     t_{tms} = \frac{t_{tm}  t_{ms} exp(i k_{mx} q) }{1+r_{tm} r_{ms} exp(2i k_{mx} q) },
\end{equation}
where $t_{ij}$ is the two-layer transmission coefficient that is related to the corresponding single-interface reflection coefficient by
\begin{equation}\label{tij}
    t_{ij} = 1+r_{ij}.
\end{equation}
The reflectivity R is the absolute square of the reflection coefficient i.e., $R = |r_{tms}|^2$, and transmission T can be calculated as $T = |t_{tms}|^2.$
Since incident light is assumed to be TM-polarized to excite SPPs, so, the electric field enhancement factor for a TM-mode is related to the magnetic field enhancement factor and can be written as \cite{reather1988surface}
\begin{equation}\label{electric field enhancement}
    T_{el} = \frac{\epsilon_t}{\epsilon_s} |t_{tms}|^2 .
\end{equation}
In the following, Eqs. (\ref{ reflection coefficient})-(\ref{electric field enhancement}), are used to observe the excitation of SPPs. 
\section{Results and Discussions}

This section contains an analysis of the proposed scheme for FWM-based amplification and excitation of SPPs and the propagation of symmetric and anti-symmetric modes for the thin silver film. First, we explore the SPP's excitation by investigating the required condition for coupler-free excitation. At the interface between the top medium (vacuum or air) with $\epsilon_t=1$, and the metal the wave vector of SPPs becomes $k_{SPP}>k_0$, where $k_{SPP}=k_0\sqrt{\frac{\epsilon_t \epsilon_m}{\epsilon_t + \epsilon_m}}$, therefore, the momentum mismatch $(\hbar k_{SPP}>\hbar k_0)$ arises that prevents the excitation of SPPs directly from light. To overcome this obstacle, an earlier approach employed a medium possessing a refractive index of $n_c$ as a coupler to increase the momentum of light to $n_c\hbar k_0$ to enable the excitation of SPPs via light.
\begin{figure}[t]
    \centering  
   \includegraphics[width=12cm]{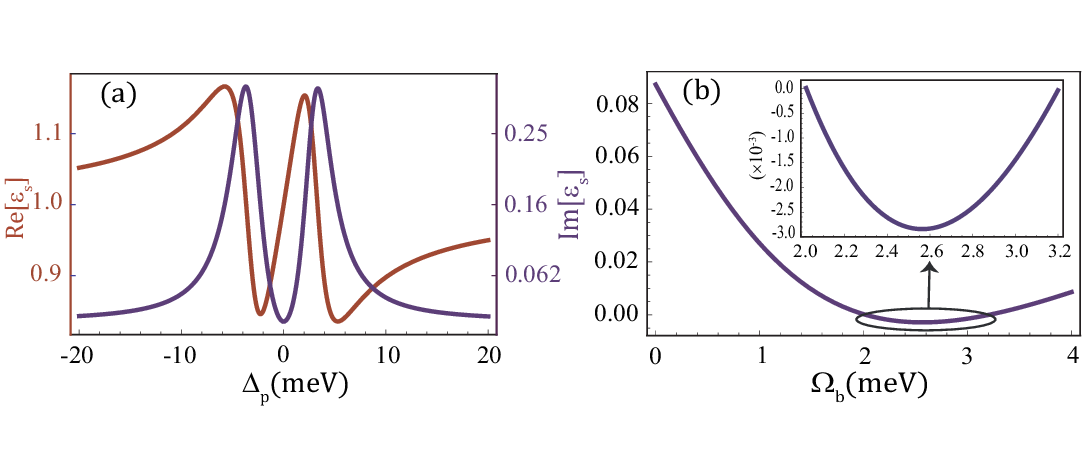}
   \caption{(Color online) (a) Real (red curve) and imaginary part (purple curve) of the permittivity of the SQWs as a function of probe detuning $\Delta_p$ (b) imaginary part of permittivity as a function of pump field $\Omega_b$ and inset shows the zoomed-in region from $\Omega_b=2$ meV to $\Omega_b=3.2$ meV. Other parameters are $\Omega_c=4$ meV, $\Omega_p=\Omega_s=1$ meV, $\gamma_2=0$, $\gamma_{3l}=\gamma_{4l}=2.07$ meV, $\gamma_{3d}=\gamma_{4d}=2.58$ meV, $\Delta_b=0$, $\Delta_c=0$, $n_t=1$, $\epsilon_t=1$, $\epsilon_m=-13.3+0.883i$, $\theta_p = 77^{\circ}$, $\lambda_0=589.1$nm and $q=50$nm.}
   \label{permittivity}
\end{figure}
However, in our proposed system, we consider the FWM-based coupler-free excitation of SPPs at the metal-SQWs interface. If the permittivity of SQWs $\epsilon_s$ becomes less than unity due to EIT effects, then it satisfies the resonance condition $k_{SPP}=k_0 n_t \text{sin}\theta_p$ for a particular angle of incidence $\theta_p$. This yields the excitation of SPPs via FWM without using any coupler.

In the absence of pump field $\Omega_b$, our system reduces to a standard three-level $\Lambda$-type configuration, which has already been discussed in atomic-vapor for coupler-free excitation of SPPs \cite{PhysRevA.91.013817}. In Fig. \ref{permittivity}(a), we plot the real and imaginary parts of the permittivity $\epsilon_s$ of SQWs as the function of probe detuning $\Delta_p$ for $\Omega_b=0$. We here choose an optimal thickness of the metal film, $q=50$, that allows the sharp SPR resonance and supports only single-interface like SPPs. The other parameters are as follow: $\Omega_c=4$ meV, $\Omega_p=\Omega_s=1$ meV, $\gamma_2=0$, $\gamma_{3l}=\gamma_{4l}=2.07$ meV, $\gamma_{3d}=\gamma_{4d}=2.58$ meV, $\Delta_b=\Delta_c=0$, $n_t=\epsilon_t=1$, $\epsilon_m=-13.3+0.883i$, $\theta_p=77^0$, $\lambda_0=589.1$ nm.  The steep normal dispersion (red curve), and the transparency window (purple curve) in the vicinity of resonance, i.e., $\Delta_p=0$, show the real and imaginary parts of SQWs permittivity $\epsilon_s$. The SPR condition, $k_{SPP}=k_0 n_t \text{sin}\theta_p$ is satisfied at $\Delta_p=-1.73$ meV for probe angle of incidence $\theta_p=77^0$, where $\text{Re}[\epsilon_s]<1$ but $\text{Im}[\epsilon_s]\neq 0$. The losses in the system, which are represented by the imaginary part of the permittivity of the SQWs $(\text{Im}[\epsilon_s]\neq0)$ reduce the sharp SPR and limit their propagation length. To compensate for the losses, we apply a third pump field $\Omega_b$ that initiates the transition between states $|2\rangle$ and $|4\rangle$, and along with the probe and control field induces a FWM process, which is responsible for the gain. Fig. \ref{permittivity}(b) illustrates the effect of the pump field $\Omega_b$ on the imaginary part of the permittivity $\epsilon_s$ for the fixed probe detuning at $\Delta_p=-1.73$ meV. Clearly, the absorption decreases with an increase in pump field, and at $\Omega_b=2$ meV, it reduces to zero. Further increase in pump field indicates gain in the system that reaches its maximum value at $\Omega_b=2.5$ meV, it then tends to decrease and retrieves zero absorption at $\Omega_b=3.2$ meV, see Fig. \ref{permittivity}(b) (the inset shows the zoomed-in (gain) region). 

\begin{figure}[t]
    \centering
    \includegraphics[width=12cm]{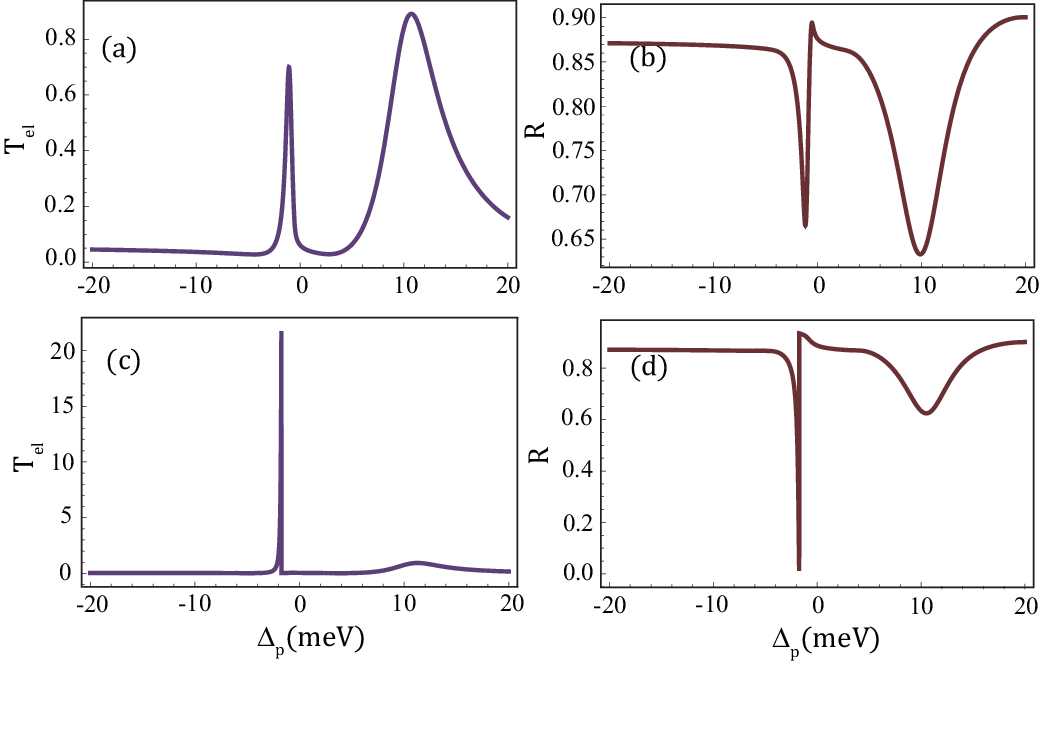}
    \caption{(Color online) Spectrum of the (a, c) field enhancement factor and (b, d) reflectivity of the probe field as a function of probe detuning $\Delta_p$ for (a, b) $\Omega_b=0$, and (c, d) for $\Omega_b=2$ meV. Other parameters are the same as in Fig. \ref{permittivity}.}
    \label{R and T}
\end{figure}
\begin{figure}[t]
    \centering
    \includegraphics[width=12cm]{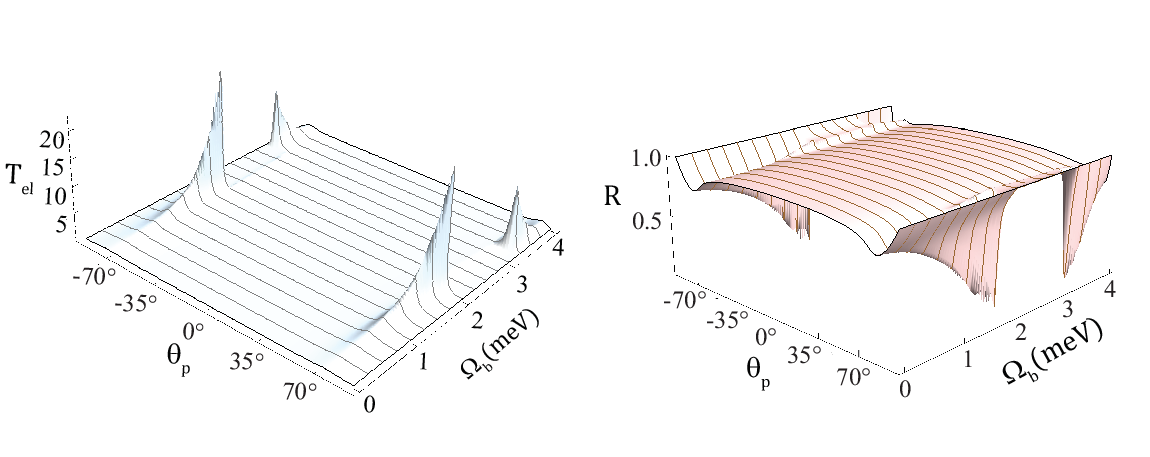}
    \caption{(Color online) 3D angle spectrum of (a) the field enhancement factor and (b) reflectivity of the probe field with respect to the pump field $\Omega_b$ and probe incident angle $\theta_p$ at fixed probe detuning $\Delta_p=-1.73$ meV. Other parameters are the same as given in Fig. \ref{permittivity}.}
    \label{angle spectrum}
\end{figure}

In Fig. \ref{R and T}, we examine the reflectivity $R$ and the field enhancement factor $T_{el}$ of the probe field with respect to the probe detuning $\Delta_p$. We also investigate the effect of the pump field on the reflectivity $R$ and the field enhancement factor $T_{el}$.  Fig. \ref{R and T} (a, b) shows the results obtained in the absence of the pump field.  In this case, the field enhancement factor is less prominent, and reflectivity is not zero, therefore, SPR effects are not so pronounced. In contrast, when $\Omega_b$ is tuned to $2$ meV, the absorption reduces to zero, and the field enhancement factor $T_{el}$ attains a sharp peak with a maximum value, and the reflectivity $R$ turns to exactly zero with a sharp dip at negative probe detuning $(\Delta_p=-1.73$ meV), indicating the resonant excitation of SPPs as shown in Fig. \ref{R and T} (c, d).

In the following, we inspect the angle spectrum of field enhancement factor $T_{el}$, and reflectivity $R$ with respect to the pump field. Owing to the pump field, significantly sharp field enhancement (see Fig. \ref{angle spectrum}(a) ) and exactly zero reflectivity (see Fig. \ref{angle spectrum}(b) ) are obtained equivalently at the points of zero absorption i.e., $\Omega_b=2$ and $\Omega_b=3.2$ meV at two different angles, called resonance angles. This shows the symmetric nature of the angle spectra about $\theta_p=0$. The weak SPR effect at $\Omega_b=0$ justifies that the only control field is not adequate to achieve sharp SPR. Interestingly, within the gain region (from 2 to 3.2 meV), there is zero field enhancement and a total reflection, as shown in Fig. \ref{angle spectrum}(a, b), respectively. In essence, excited SPPs radiate into the top layer and absorb by the metal. In this case, the energy from the gain medium is transferred to the SPPs which is then radiated into the reflected wave, see Ref. \cite{jha2013quantum} for details. 

\begin{figure}[b]
    \centering
    \includegraphics[width=12cm]{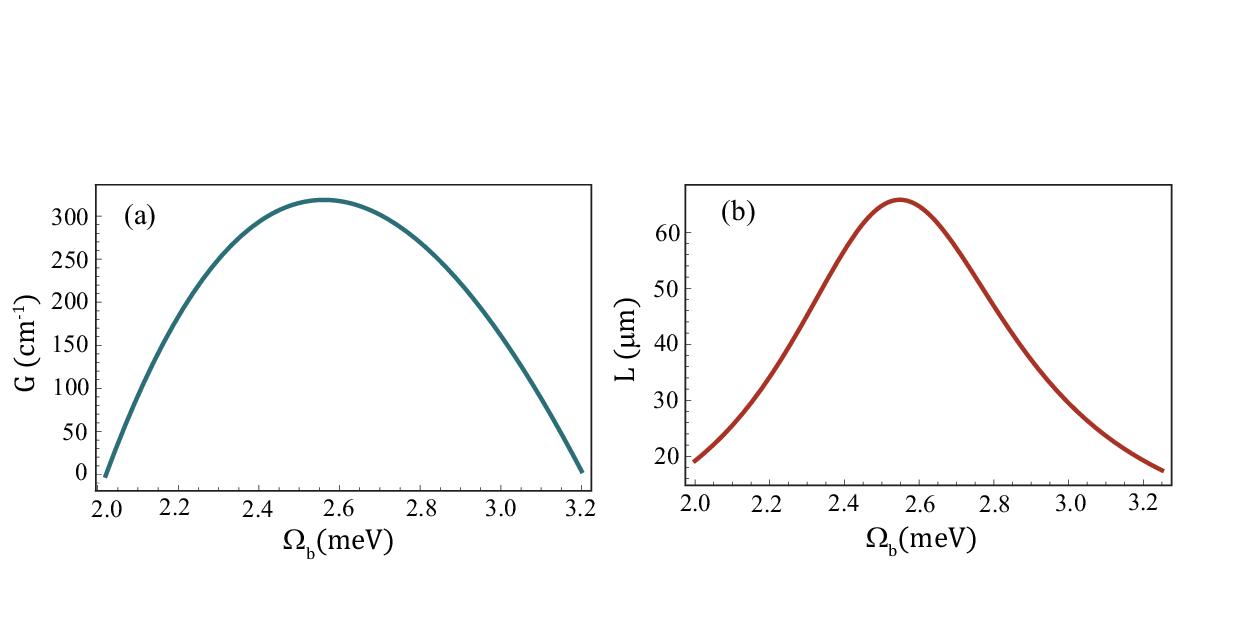}
    \caption{(Color online) (a) Optical gain power $G$ (blue curve) and (b) the propagation length (red curve) of SPPs versus pump field $\Omega_b$. Other parameters are given in Fig. \ref{permittivity}}
    \label{gain}
\end{figure}

Next, we are interested in observing the propagation length of SPPs by virtue of the gain induced by a FWM process. Essentially, propagation length is the distance traveled by SPPs before the decrease in their intensity by 1/e of the original value. It is the reciprocal of the imaginary part of the SPP wave vector, i.e., $L=1/2 \text{Im}[k_{SPP}]$. Small propagation length remains the major challenge for the practical applications of SPPs for years. In our proposed scheme, the gain due to the generated fourth wave is one of the factors to enhance propagation length significantly. The optical gain power is defined as $G=-k_0 \text{Im}[\epsilon_s]/\text{Re}[\epsilon_s]$ \cite{Nezhad:04}. In Fig. \ref{gain}, we plot (a) the optical gain and (b) the propagation length of SPPs as a function of the pump field $\Omega_b$, which is a controllable parameter for the gain. The gain increases gradually with an increase in pump field and eventually, it saturates at $\Omega_b=2.5$ meV, as shown in Fig. \ref{gain}(a). Initially, at zero absorption, propagation length is only $20$ $\mu$m. However, at the maximum value of the gain, losses are suppressed, thereby propagation length is enhanced to $60$ $\mu$m, see Fig \ref{gain}(b).

\begin{figure}[t]
   \centering
    \includegraphics[width=12cm]{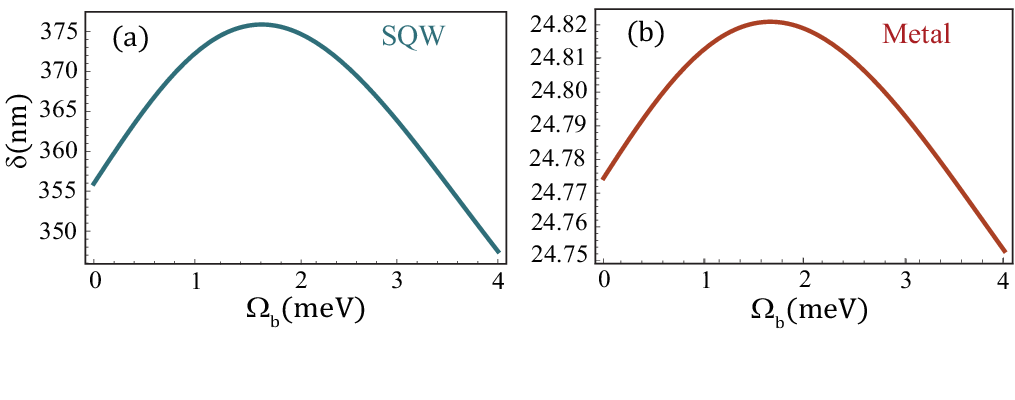}
   \caption{(Color online) Penetration depth of SPPs in (a) SQW and in (b) metal film as a function of the pump field $\Omega_b$ at fixed probe detuning $\Delta_p=-1.73$ meV. Other parameters are the same as in Fig. \ref{permittivity}}
    \label{penetration depth}  
\end{figure}

Since SPPs have evanescent fields associated with them that penetrate inside the surrounding media. The penetration depth is actually the distance that the field can penetrate inside the metal and dielectric layer and is defined by $\delta_j=\frac{1}{k_0}\sqrt{|\frac{\epsilon_s+\epsilon_d}{-\epsilon_j}|}$ \cite{barnes2006surface}, where $j=m,s$ represents metal and SQW, respectively. The penetration depth of SPPs both in SQW and metal-medium is plotted as a function of the pump field in Fig. \ref{penetration depth}(a) and (b), respectively. It is noteworthy that the field concentration in SQW is greater than that in metal. A larger penetration depth in the dielectric medium (SQW) is appreciated due to the lower rate of losses associated with them, which corresponds to a larger propagation length. On the contrary, the SPPs energy absorbed by the metal is considered as a loss due to the lossy nature of the metal. There is a characteristic trade-off between the high confinement of the SPPs at the interface and their propagation length. Therefore, it is crucial to have the optimized penetration depth for the large propagation length of SPPs and for their strong confinement at the interface. In our proposed scheme, the penetration depths of the SPPs field inside both media, as shown in Fig. \ref{penetration depth}, indicate the high field concentration at the interface.
\begin{figure}[t]
    \centering
    \includegraphics[width=12cm]{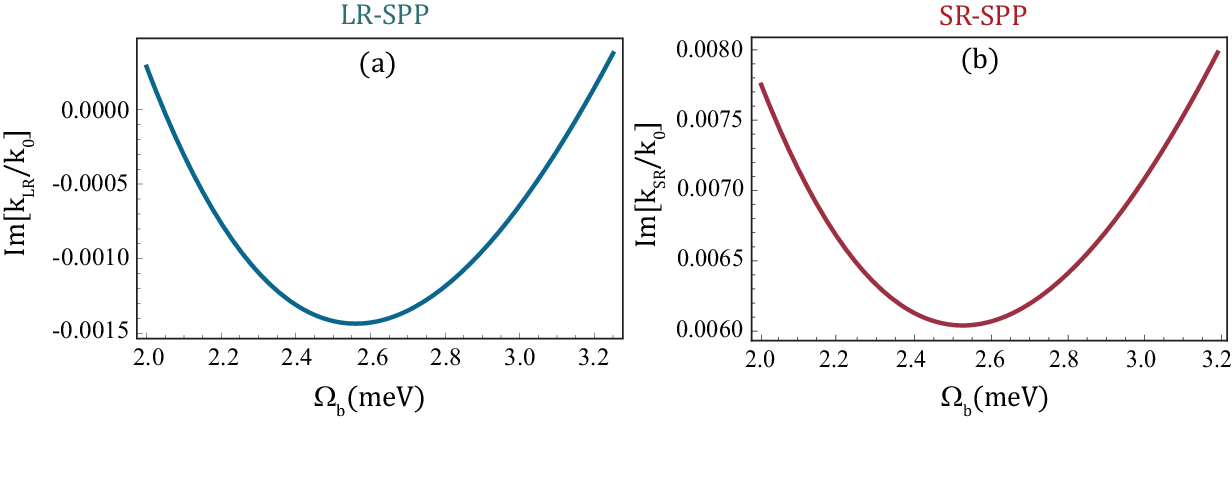}
    \caption{(Color online) The normalized imaginary part of the dispersion relation of the (a) LR-SPPs and (b) SR-SPPs against the pump field $\Omega_b$ at fixed probe detuning $\Delta_p=-1.73$ meV.}
    \label{dispersion length}
\end{figure}
\subsection{Long-range and Short-range SPPs}
For the symmetric structure presented in Fig. 1, where the permittivity of both the top-medium and SQW is comparable, and the thickness of film $q$ is much smaller than the optical wavelength $\lambda_0$, i.e., $d<<\lambda_0$, the probe field gives rise to symmetric (LR-SPP) and anti-symmetric (SR-SPP) modes of SPPs. 

Using the appropriate boundary conditions for the aforementioned symmetry, the following dispersion relation for LR-SPP is obtained \cite{raether1989surface}.
\begin{equation}\label{LR dispersion}
    \text{tanh}(k_{mx}q/2)=-\frac{\epsilon_m k_{dx}}{\epsilon_s k_{mx}}.
\end{equation} 
Similarly for the anti-symmetric mode (SR-SPP), it is given by
\begin{equation}\label{SR dispersion}
    \text{coth}(k_{mx}q/2)=-\frac{\epsilon_m k_{dx}}{\epsilon_s k_{mx}},
\end{equation}
where $k_{ix}=\sqrt{k_{j}^2-\epsilon_i(\omega /c)^2}$, for $i=m,s$, and $j=LR,SR$ that represent LR and SR-SPPs, respectively. Since $k_{j}$ cannot be presented explicitly. However, in the limit of sufficiently thin films $(q<40$ nm$)$ \cite{han2012radiation}, the dispersion relation can be simplified using the small angle approximation, i.e., $\text{tanh}x\approx x$. This gives the following explicit forms for LR-SPP and SR-SPP \cite{han2012radiation}
\begin{equation}\label{small q LR}
    k_{LR} \approx k_0 \sqrt{\epsilon_s + (k_0 \epsilon_s q/2)^2 \dot [1-(\epsilon_s/\epsilon_m)]^2},
\end{equation}
\begin{equation}\label{small q SR}
     k_{SR} \approx k_0 \sqrt{\epsilon_s +[2\epsilon_s/q k_0 \epsilon_m]^2}.
\end{equation}
The Eqs. (\ref{small q LR} and \ref{small q SR}) are utilized to exploit the several properties of LR-SPPs and SR-SPPs.
\begin{figure}[t]
    \centering
    \includegraphics[width=12cm]{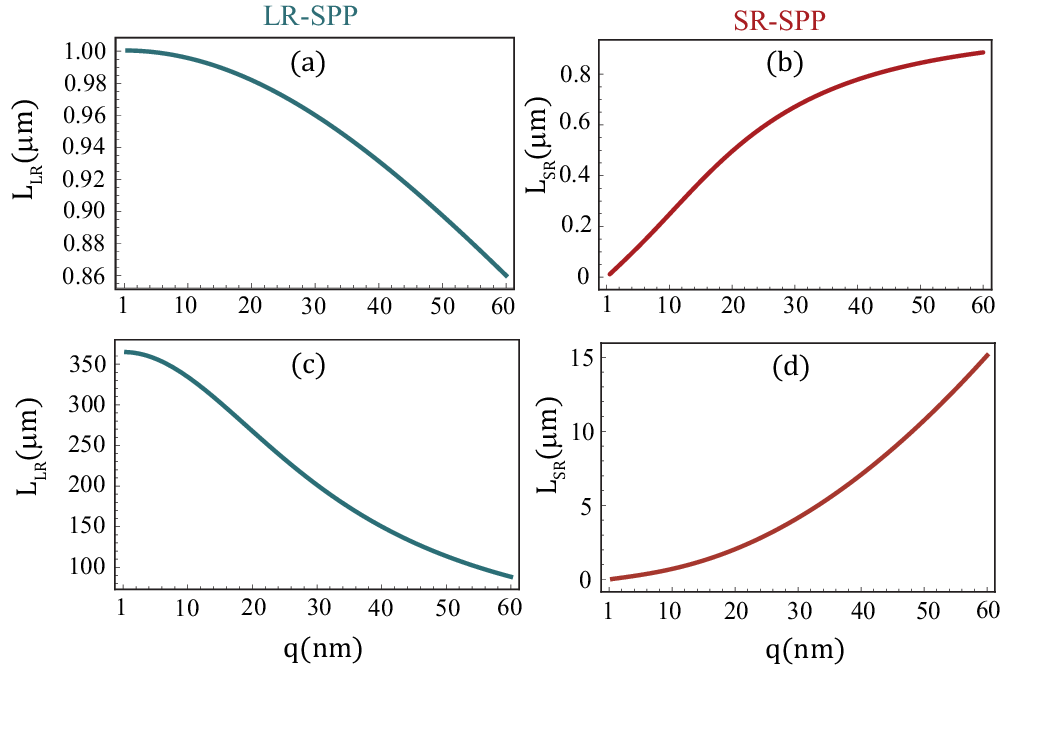}
    \caption{(Color online) Propagation length of (a,c) LR-SPPs and (b,d) SR-SPPs versus the metal film thickness $q$ for (a,b) $\Omega_b=0$ and (c,d) $\Omega_b=2$ meV. Other parameters are the same as in Fig. \ref{permittivity}}
    \label{propagation length}
\end{figure}

We start by analyzing the propagation of SPPs along a thin silver film at a certain angle of incidence. The losses experienced by SPPs are strongly linked to the thickness of the metal film \cite{reather1988surface}, therefore, it influences their excitation and propagation length. Although the presence of the gain has significantly enhanced the propagation length of SPPs. However, a two-interface structure with a thin metal film surrounded by air and a semiconductor quantum medium supports the LR-SPP that propagates to a relatively larger distance and survives for a longer period. Therefore, to observe such SPPs in our proposed scheme, we reduce the thickness of the metal film to $36.8$ nm. This is the required minimum thickness for our system at which the condition for coupler-free excitation $(k_{LR/SR}<k_0)$ holds for the LR-SPPs and SR-SPPs. Since the penetration depth in the metal medium is less than its thickness, so, the SPPs propagating at the air-metal and metal-SQW interface begin to overlap giving rise to LR and SR-SPPs. 

In Fig. \ref{dispersion length}, we plot the imaginary part of the dispersion relation given in Eq. (\ref{LR dispersion}) and (\ref{SR dispersion}) as a function of the pump field $\Omega_b$ for LR-SPP and SR-SPP. We fix thickness at $q=36.8$ nm while keeping all the other parameters the same as given in Fig. \ref{permittivity}. Evidently, around $\Omega_b=2$ meV, the absorption becomes zero and further increment in the pump field indicates gain for LR-SPPs with its maximum value at $\Omega_b\approx2.5$ meV, shown in Fig. \ref{dispersion length}(a). However, the absorption approaches a minimum value, but it never reduces to zero for short-range SPPs, see Fig. \ref{dispersion length}(b), which indicates the propagation losses experienced by the short-range SPPs.

\begin{figure}[t]
    \centering
    \includegraphics[width=12cm]{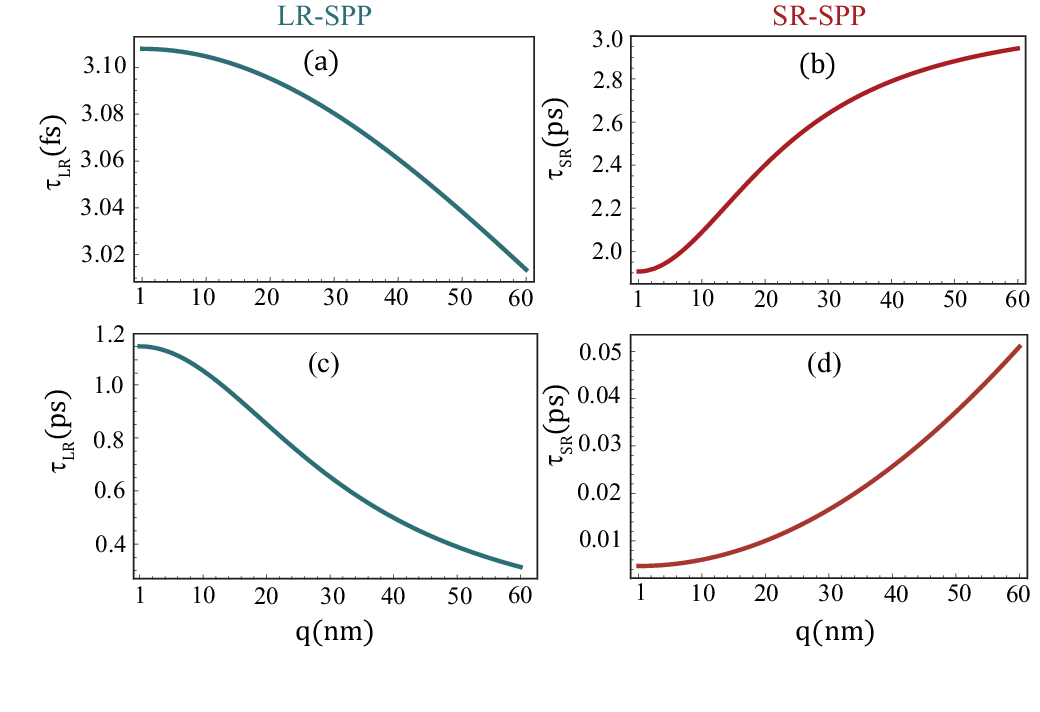}
    \caption{(Color online) The lifetime of (a,c) LR-SPPs and (b,d) SR-SPPs versus the metal film thickness $q$ for (a,b) $\Omega_b=0$ and (c,d) $\Omega_b=2$ meV. Other parameters are given in Fig. \ref{permittivity}.}
    \label{lifetime}
\end{figure}

Next, in Fig .\ref{propagation length}, we plot the propagation length ($L_{LR(SR)}=1/2 \text{Im}[k_{LR(SR)}]$) of both LR and SR-SPPs for two different cases: (a) $\Omega_b=0$, and (b) $\Omega_b=2$ meV. For a smaller thickness of the silver film, less energy is localized inside the metal film. This results in less confinement at the interface, and therefore less dissipation and a larger propagation length for LR-SPPs, as shown in Fig. \ref{propagation length}(a). 
Whereas, reducing the film thickness for short-range SPPs results in a larger energy distribution inside the metal film, and thus high concentration at the interface. This further leads to strong dissipation and their small propagation length, see Fig. \ref{propagation length}(b). In the absence of the pump field, the overall propagation length is small enough even for LR-SPPs. However, the presence of the pump field enhances the propagation length significantly to $350$ $\mu$m for LR-SPPs, see Fig. \ref{propagation length} (c) which is far larger than the SPPs propagating at thick metal-SQW interface (shown in Fig. \ref{gain}(b)). The maximum concentration of LR-SPPs in the SQW (dielectric) medium is actually responsible for their larger propagation length. On the contrary, the propagation length of SR-SPPs is still shorter even in the presence of the pump field as shown in Fig. \ref{propagation length}(d).

Subsequently, the effect of the metal thickness $q$ is investigated on the lifetime of LR-SPPs and SR-SPPs. The lifetime is the duration that SPPs persist before their absorption into the surrounding medium. It is defined as $\tau_{LR(SR)}=L_{LR(SR)}/v_g$, where $v_g$ is the group velocity which is obtained as $[\frac{\partial k_{LR(SR)}}{\partial \omega}]^{-1}$. It is important to emphasize that the increased lifetime is essential to the effective slowdown of SPPs. In Fig. \ref{lifetime}, we analyze the effect of the pump field on the lifetime of LR and SR-SPPs. Fig. \ref{lifetime}(a, b) shows the lifetime of LR-SPPs and SR-SPPs in the absence of the pump field. Interestingly, the lifetime behaves in a similar manner as does the propagation length for both modes. The high concentration of these waves on the interface increases their coupling strength with light leading to a high energy dissipation rate. Therefore, in the absence of the pump field, LR-SPPs have a lifetime of femtoseconds and it is maximum for extremely small thicknesses. Conversely, a small thickness corresponds to a shorter lifetime for SR-SPPs. However, as shown in Fig. \ref{lifetime}(c), the presence of the pump field increases the lifetime for LR-SPPs from femtoseconds to pico-seconds and is much longer than the lifetime of SR-SPPs (see Fig. \ref{lifetime}(d)). Hence, the FWM process plays a crucial role to sustain the SPPs for a relatively longer period with their extended propagation distance for sufficiently thin films. Based on the different propagation lengths and the concentration at the interface, LR-SPP and SR-SPP can be utilized for various purposes in practical applications.
\section{Conclusion}
In this paper, we have proposed a scheme to observe the amplification of SPPs with their enhanced propagation length using a four-level asymmetric semiconductor quantum well structure that exhibits the FWM process. We have also investigated the excitation of SPPs through quantum interference effects by using optimally thick metal film. In our system, the simultaneous interaction between the applied three electromagnetic fields generated a fourth wave through the EIT-based FWM process, which played a significant role to compensate the losses. As a result, sharp SPR resonance and significantly enhanced propagation length of SPPs are observed. In addition, the sensitivity of the SPPs to the probe detuning and angle of incidence is also investigated. Further, we have explored the formation of two distinct SPPs modes for the thin metal film in our proposed scheme. It is figured out that the propagation length and lifetime of these modes are significantly enhanced at zero absorption due to the FWM process. Evidently, the excitation and amplification of SPPs along the metal-SQW interface in the visible region are achieved. Our system also supports the LR-SPPs with their relatively larger propagation length and longer lifetime.
\begin{backmatter}
\bmsection{Disclosures}
The authors declare no conflicts of interest.

\bmsection{Data availability} Data underlying the results presented in this paper are not publicly available at this time but may be obtained from the authors upon reasonable request.

\end{backmatter}
\newpage
\bibliography{bibliography.bib}

\end{document}